# Structural Control of Atomic Silicon Wires

Furkan M. Altincicek[1], Christopher C. Leon[1,4], Lucian Livadaru[1], Taras Chutora[1], Max Yuan[1], Roshan Achal[2], Jason Pitters[3], Robert Wolkow[1,2]*

[1] *Department of Physics, University of Alberta, Edmonton, Alberta T6G 2E1, Canada*

[2] *Quantum Silicon Inc., Edmonton, Alberta T6G 2M9, Canada*

[3] *National Research Council of Canada, Edmonton, Alberta T6G 2M9, Canada*

[4] *Current address: Département de chimie, Université Laval, Québec, Québec G1V 0A6, Canada*

* Correspondence to: rwolkow@ualberta.ca

## Abstract

Bare Si(100)-2×1 surface atoms exhibit a buckled structure where one Si atom in a dimer is lowered while the other is raised, leading to two possible buckling configurations equivalent in energy. The relatively low energy barrier between these configurations allows dimers to flip rapidly and uncontrollably unless stabilized by surface defects or observed at low temperatures due to reduced thermal energy using scanning tunneling microscopy (STM). This rapid flipping results in a time-averaged symmetric appearance under STM. In this study, we investigated variable length buckled dimer wires on the hydrogenated Si(100) surface composed of silicon dangling bonds for the first time. We demonstrate that on degenerate p-type silicon at 4.5 K, the rapid switching of these dimers can be frozen at low scanning biases. Such buckled wires can however be controllably flipped using a bias pulse. A line as long as 37 dimers was repeatedly uniformly flipped by a single pulse delivered near one terminus of the wire. The tip-directed flipping of a particular wire does not switch adjacent wires, suggesting binary wires can make well isolated rewritable binary memory elements. Furthermore, at sufficiently high biases switching generates telegraph noise that could be of utility for random number generation. The integration and encapsulation of these wires with previously described silicon dangling bond-made logic gates and binary wires might allow for self contained actuation and readout without requiring any role of an STM tip.

## Introduction

The silicon (100) surface has garnered significant interest for decades due to its crucial role in semiconductor technology.[1, 2] The Si(100)-2×1 reconstructed surface has been extensively studied.[1-15] This structure features bonded pairs of neighboring silicon atoms, known as dimers, which periodically repeat to form rows. Dimerized Si atoms exhibit one dangling bond per atom combining into collective states that span the length of the line, while retaining some of the dimer $\pi$ bond character. On the bare surface, dimers undergo buckling. In a buckled dimer one Si atom substantially transfers an electron to the other, resulting in one $sp^3$-like atom being raised with respect to the surface plane (negatively charged), while the other becomes $sp^2$-like and is lowered within the same dimer (positively charged).[2, 5] This buckling pattern alternates for successive dimers in a row (Figure 1e). Although buckling images were captured in the early years of scanning tunneling microscopy (STM)[6, 7], this phenomenon may not always be visible due to the thermal energy of the system or the tunneling current applied by the STM tip.[8, 9] These factors cause dimers to switch rapidly between the two buckled orientations, leading to a symmetric averaged appearance. Buckling becomes more readily observable under STM when a surface defect pins one of the silicon atoms, initiating buckling along that dimer row.[10, 11] Additionally, at lower temperatures, dimers are more likely to appear buckled due to reduced thermal energy, which diminishes the ability to overcome the energy barrier for flipping the buckled orientation.[12, 13] Due to these limitations, so far only theoretical studies on the two buckled configurations of dimers have proposed their usage as memory elements and logic gates.[16-18]

The hydrogen-terminated Si(100) surface also exhibits a 2×1 unit cell, where dimers consist of two surface hydrogen atoms attached to underlying silicon dimers. With the silicon atoms on the surface terminated by a hydrogen monolayer, there is no longer the potential for charge transfer within a dimer. Consequently, hydrogen-terminated silicon dimers do not exhibit buckling and remain symmetric (parallel to the surface).

The hydrogen monolayer can serve as a resist, allowing individual hydrogen atoms to be removed with an STM tip, leaving behind silicon dangling bonds (DBs),[19-21] which have electronic states within the silicon band gap. Diverse applications for nanoelectronics have been proposed.[22] The controlled creation of individual DBs or pairs of DBs is achieved by applying bias pulses from the STM tip, which breaks the hydrogen-silicon bond. It has been shown that in this process the removed hydrogen atom can adhere to the tip, allowing it to be controllably redeposited onto a DB, enabling editable hydrogen lithography.[23-25] These advancements have led to numerous studies demonstrating the significance and application of DBs.[25-32]

The removal of pairs of hydrogen atoms along the dimer rows leaves bare dimers on the hydrogenated surface, referred to as dimer wires in this paper. STM scanning parameters for this surface are typically set above +1 V or below -1 V to avoid scanning too closely to the surface or within the bandgap, which could result in an STM tip contact or the destruction of the custom-made DB patterns on the surface. Scanning dimer wires at these voltages causes the exposed dimer wires to flip rapidly between the two buckled configurations, making them appear symmetric even at low temperatures due to current-induced flipping.[9, 33, 34]

Dimer wires' filled states appear symmetric (Figure 1a) under standard STM scanning parameters (more negative than ~-0.3 V) due to the rapid switching of dimers, which occurs faster than what the STM can capture due to bandwidth limitations.[33, 34] The switching process is induced by the current from the tip which has the effect of heating the surface locally.[19,35] At reduced negative bias, the process slows enough for the STM to capture the random switching (Figures 1b,c). Finally, at sufficiently low bias, the process slows to a point where the vibrational modes have enough time to relax, allowing the fixed buckled structure to be imaged by the STM (Figure 1d).

It is important to note that the change from one buckled state of a single dimer to the opposite is not a normal vibration mode. The transition between the two buckled states is an activated process with the symmetric dimer structure being the transition (barrier) state.

In this study, we investigated variable-length buckled dimer wires on a hydrogenated, degenerately doped p-type silicon sample (~0.005 Ω·cm). At low negative biases, dimer wires can be observed in buckled states at 4.5 K. Furthermore, we demonstrated that the buckling orientations can be controllably flipped using bias pulses. Our results show that switching one wire this way, even when separated from another wire by just a single row of hydrogenated dimers, does not cause the adjacent wire to switch. Finally, we present Scanning Tunneling Spectroscopy (STS) results on individual wires, revealing telegraph noise in a regime where the switching rate is neither too high nor too small, which we further analyzed through statistical methods.

## Results and Discussions

Figure 1 presents a series of constant-height dI/dV maps of three bare dimers (six dangling bonds in a 2×3 array) that visualizes the process described above. The imaging bias values are -300 mV, -250 mV, -225 mV, and -200 mV for Figures 1a-d, respectively. At a constant tip-sample distance corresponding to -200 mV and 10 pA on an adjacent area of hydrogen-terminated Si, reducing the bias decreases the tunneling current, thereby slowing the flip-flop motion of buckled dimers. This is evident from the color bar ranges in the subfigures, where the maximum current values are 373 pA, 193 pA, 142 pA, and 90.5 pA, respectively. The state density shown in Figure 1a is narrower than the wire's physical dimensions, indicating rapid switching. Due to this faster switching rate, which exceeds the STM's capture capability, we observe a time-averaged appearance of the two buckled configurations converging along the centerline of the dimer wire. As the bias is reduced, as seen in Figure 1b, buckling becomes observable, but random switching occurs, resulting in noisy-appearing lines. Figure 1c shows the threshold for uncontrollable buckling switching, with a single switching event visible as a streak. Finally, Figure 1d depicts the complete elimination of uncontrollable buckling switching, resulting in a stationary image corresponding to a single geometry. On this degenerately doped p-type silicon sample, the Fermi level is near the valence band edge, allowing imaging of the surface and wires at small negative biases. A ball and stick model of the buckled 3-dimer wire captured in Figure 1d, is given in Figure 1e, in which the symmetric surface hydrogen terminated dimers are also visible.

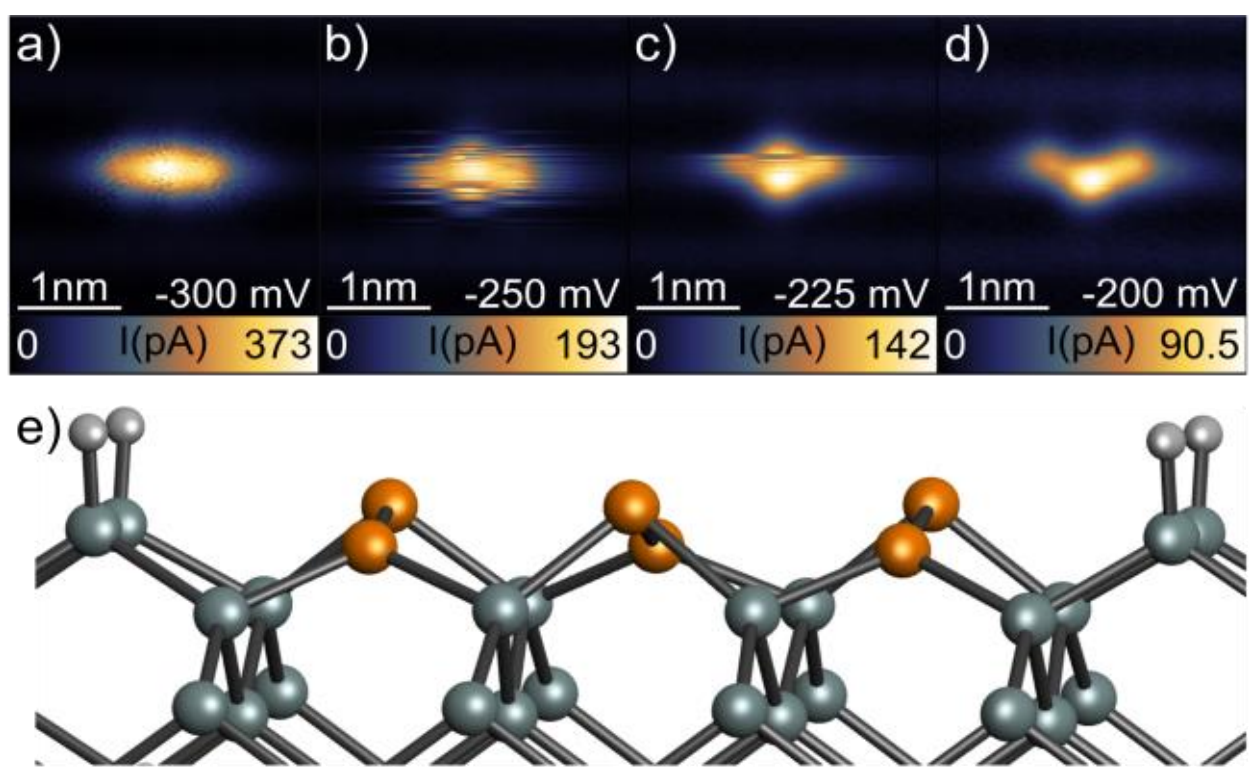

**Figure 1)** Constant height dI/dV images of 3 bare dimers taken at **a)** -300 mV, **b)** -250 mV, **c)** -225 mV, and **d)** -200 mV. The series illustrates the decrease in the flipping rate from a) to d) as the bias and therefore the current is reduced. **e)** Side view of the ball and stick model illustrating the static wire structure observed in d), where Si atoms are gray (and orange for the surface atoms) and H atoms white; the vertical is the [100] direction of the crystal. The height setpoint was -200 mV and 10 pA on hydrogen-terminated Si.

Once the pre-made dimer wires are captured in their buckled states at low biases, their orientation can be flipped by briefly increasing the bias to induce inelastic tunneling, effectively providing a burst of heat to overcome the barrier. In this study, we applied bias pulses of -300 mV for 50 ms to raised DBs to cause controlled switching. The bias pulse provides sufficient energy as inelastic tunneling electrons causing a local heating effect to overcome the flipping barrier without desorbing nearby hydrogen atoms. This process is illustrated in Figure 2. Figure 2a shows the initial configuration of two bare dimers. Figure 2b presents a scan paused mid-scan, with the rastering direction from bottom to top. During the pause, the tip was moved over the raised DB on the left, and a bias pulse was applied, causing the orientation to switch and creating a line artifact similar to the one shown earlier in Figure 1c but this time intentionally induced. Figure 2c depicts the aftermath, where the heights of all four DBs have shifted, and the orientation has flipped. Due to the said scanning direction, the bottom half of Figure 2b corresponds to Figure 2a, while the upper half matches Figure 2c, continuing from after the pulse. To demonstrate that these wires can be made as long as desired and that the switching mechanism remains effective, we show a 37-dimer-long wire in Figures 2d and 2e in two different configurations that were switched under control by a pulse applied over a terminal dimer.

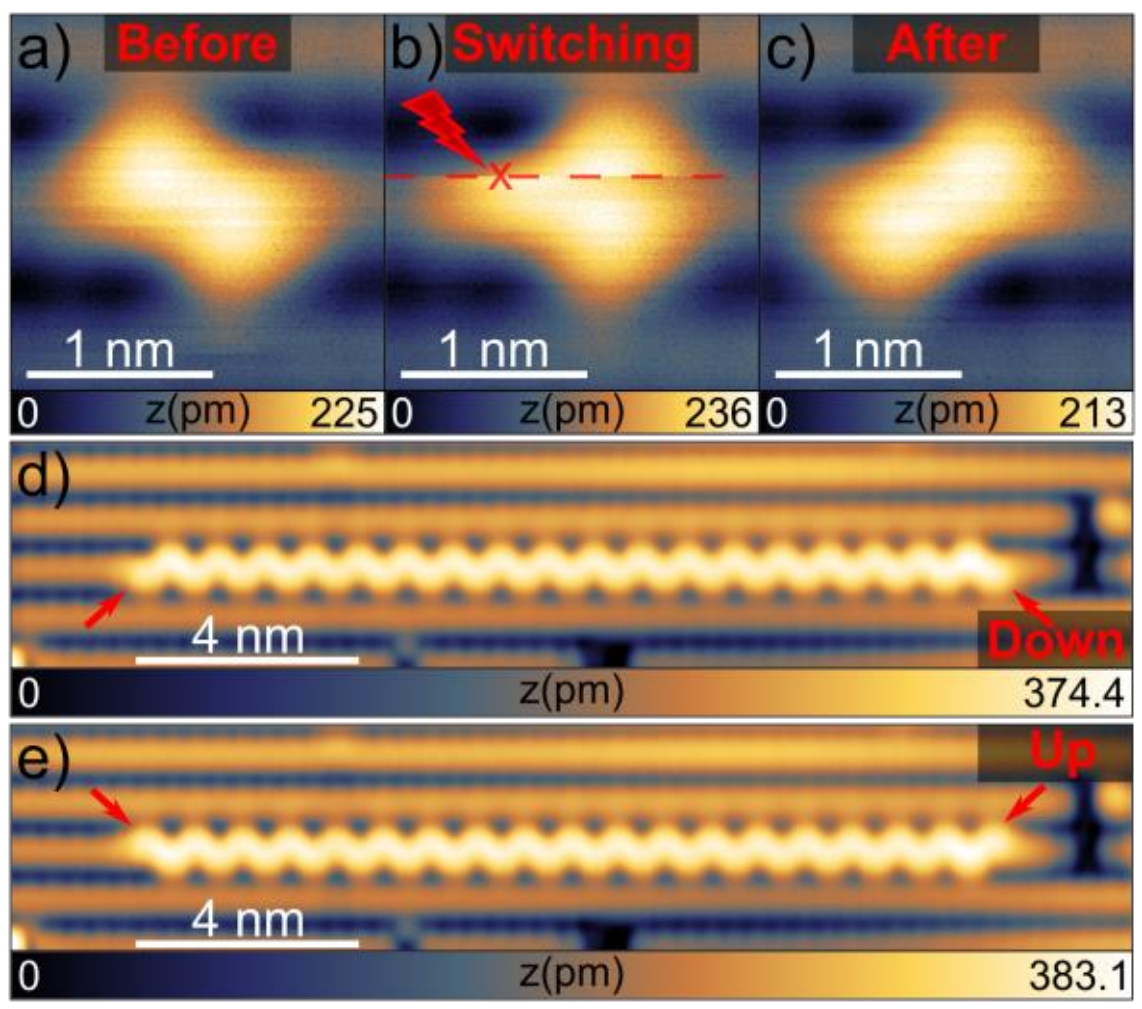

**Figure 2)** Constant-current STM images taken at -300 mV with 5 pA current, showing **a)** the initial configuration, **b)** during, and **c)** after the controlled switching of 2 bare dimers via a bias pulse. The STM tip scans along horizontal lines which are rastered from bottom to top. The x mark on the dashed red line in b) indicates the location where the STM pulse was applied and geometry switched. **d)** A 37-dimer wire in the "Down" orientation, and **e)** the same wire in the "Up" orientation after controlled switching, imaged at a constant current of 50 pA at -300 mV.

As previously discussed in the literature,[35] the most common flipping mechanism for a dimer wire is by means of a migrating geometric kink (a misalignment in the otherwise perfectly alternating buckled geometry pattern) propagating along the length of the wire. This was also called a phason in the above reference and it is similar to a domain wall in 1D, separating wire segments of opposite buckling pattern (or phase). This flipping mechanism is consistent with our experimental observation that the pulse only needs to be applied at one location in order to flip the entire wire.

In order to further ascertain and quantify this flipping mechanism, we undertook a series of simulations by the density functional theory (DFT) method. We simulated the ground state properties of dimer wires of various lengths by performing DFT geometry optimization calculations with the GGA(PBE) exchange correlation functional,[36] using the AMS 2024 program.[37-42] Model dimer wires were constructed in a periodic supercell (slab) model having 3 dimer rows, 12 dimers each at the surface, and 6 layers in depth (see Figure S1a). More details are found in the Supplemental Information section [SI].

First we verified by geometry optimization (Figure S2) that the ground state of a dimer wire consists of a zig-zag of alternating bucked dimer bonds, and is doubly degenerate. We looked at relatively small wire lengths: 3, 4, 5, and 7 dimers. We then calculated, by the nudged elastic band (NEB) method,[43-45] the transition from the ground state configuration to kinked configurations of higher energy, which enable the flipping of the whole wire configuration. NEB calculations were performed using smaller models for efficiency (Figure S1b and S1c) and details are given in the Supplemental Information section [SI].

We performed NEB calculations for wire lengths 3, 5, and 7. The transition between the ground state and end kink states is very similar in all cases. The geometry and total energy of initial, intermediate, and final states are obtained from each NEB calculation. The transition (barrier) state in each case is one in which the end dimer bond is flat (surface parallel). The barrier height from the ground state is 270, 255, and 261 meV for the three cases, respectively. From the kink state, the barrier is 120, 130, 117 meV, respectively.

As an example, in Figure 3, we show our results for the transitions between the lowest energy configurations of a 5-dimer wire. Two NEB calculations (using 8 images each) are joined here to illustrate kink propagation along the wire: the 1st NEB run is between the ground state configuration and the state with a kink at the end of the wire; the 2nd NEB run is between this latter state and the state with the kink migrated toward the interior of the wire by one dimer. The second barrier, 166 meV is significantly lower than the first, 255 meV, and the first inner kink state is lower than the end kink state by 47 meV.

Supported by these simulations and previous studies,[35] the most common mechanism for the whole wire flipping is: hot electrons tunneling from the STM tip raise the local temperature associated with the vibration of the end dimer of the wire, aided by a low de-excitation rate of the dimer rocking mode. When the STM bias is sufficiently high to provide enough energy via hot electrons (> 260-270 meV) to overcome the barrier from the ground state to the nearest local minimum state, a kink will form at the end of the wire. This metastable, local minimum state is separated by a 170-210 meV barrier from the other local minima (kinks at other nearby locations). Subsequent hot electrons will again induce a local temperature increase and the kink will shift (migrate) along the wire until it anneals at one end. When a kink is formed at one end and is annealed at the other end, the entire wire has been geometrically flipped.

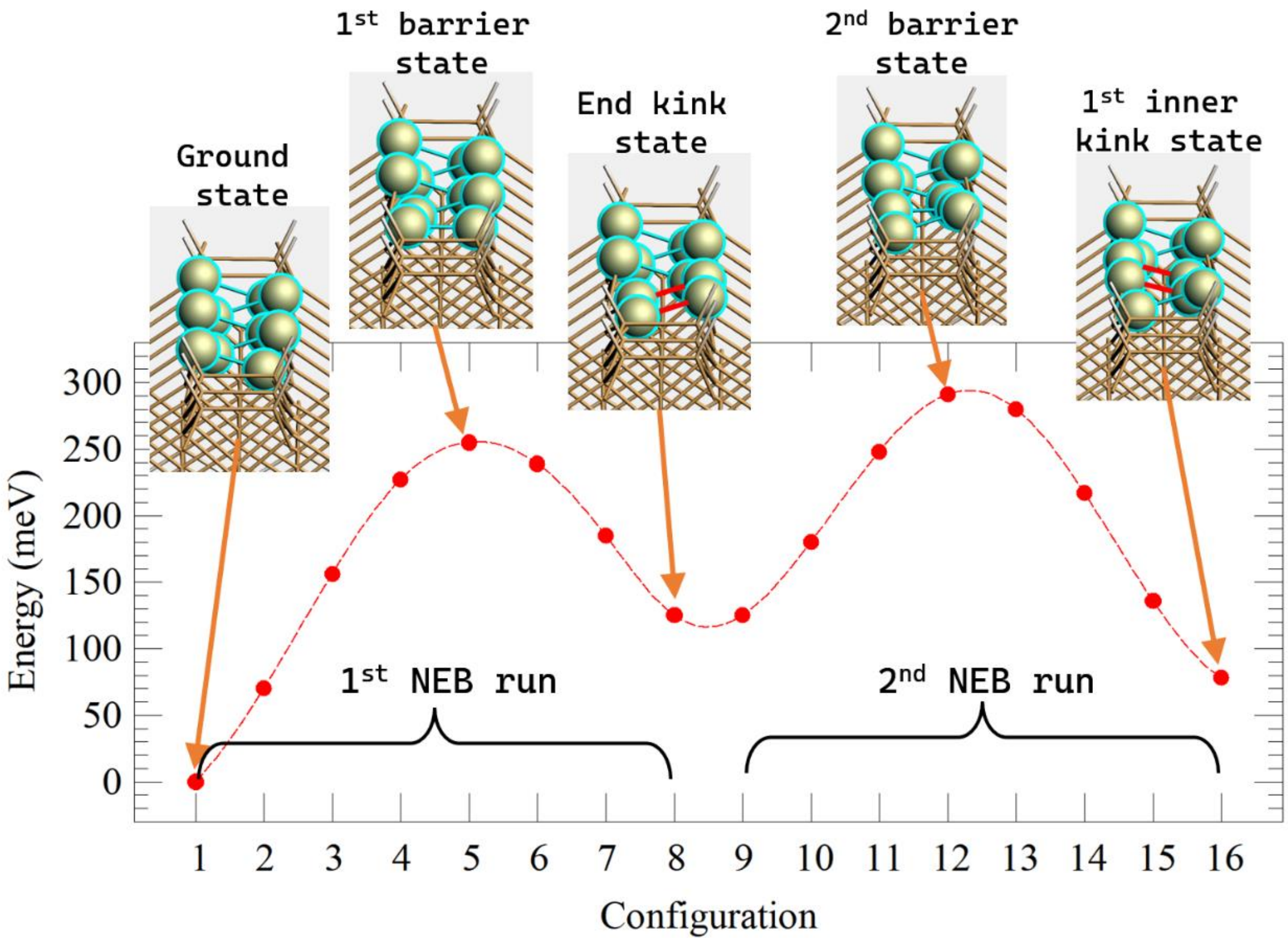

**Figure 3)** Kink formation and migration. Results of NEB calculations on the transitions between the lowest energy configurations of a 5-dimer wire. Two NEB calculations (using 8 images or configurations each) are joined here to indicate kink propagation along the wire. The first NEB calculation is for the transition from the GS to the state with a kink at the very end. The second NEB calculation is for kink displacement by one dimer. Local minima and barrier states are also illustrated pictorially using a molecular model with only the Si dimers shown in green as ball-and-stick models. Kink states are marked by red colored parallel bonds at the location of the kink.

To explore the potential applications of multiple dimer wires, we investigated dimer wires separated by a single row of hydrogen terminated dimers. Figure 4 presents a series of constant-current STM images showing two 9-atom-long dimer wires, each separated by a row of hydrogen dimers. Figures 4a-c display the bottom wire (indicated by the red box) induced to flip back and forth while the top wire (indicated by the green box) maintains its orientation. Conversely, Figures 4c-e show the top wire being flipped back and forth while the bottom wire remains unchanged. This demonstrates that there is insufficient crosstalk between the wires to cause an adjacent wire to flip while a targeted wire is acted upon. The independent switching of the two nanostructures, separated by approximately 1.2 nm, can be attributed to two main factors. First, DBs are gap states and therefore are substantially electronically decoupled from the bulk, providing no direct pathways for electrons to travel to nearby structures. Second, the tunneling current depends exponentially on the tip-sample distance, so a bias pulse applied to one structure with a sharp STM tip has minimal chance of perturbing the other structure. This indicates that the switching mechanism is dynamically induced by an electronic transition (DIET) rather than by an electric field (DIEF).[46,47] The capability to flip the entire structure demonstrates a potential use as atomic-scale memory elements, where two buckling orientations can represent binary 0 and 1 bits. Additionally, the length of the buckled wires introduces an extra degree of information.

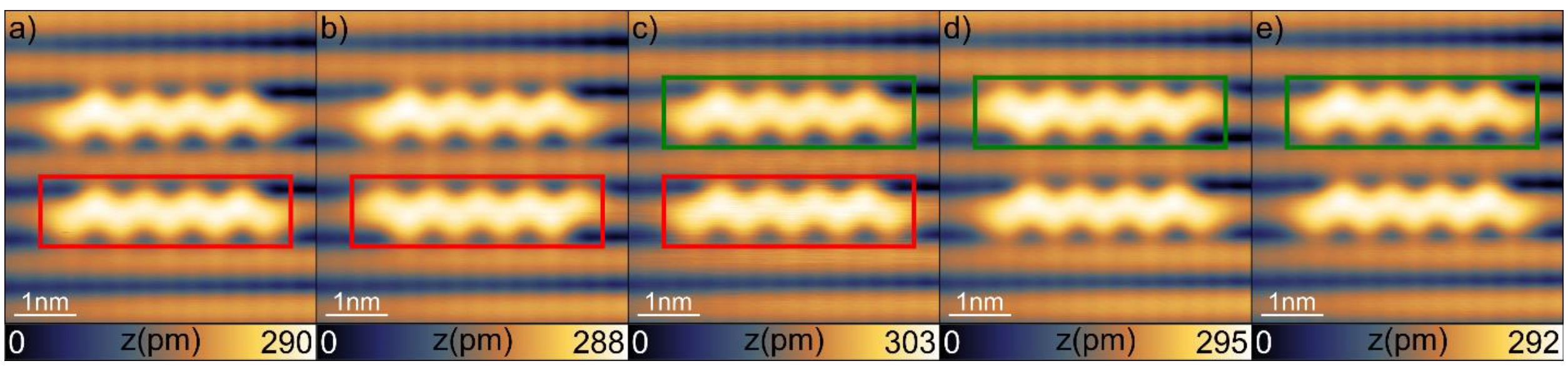

**Figure 4)** Series of constant-current STM images taken at -350 mV with 5 pA current, demonstrating no crosstalk between two 9-atom-long dimer wires separated by a single row of hydrogen dimers. In **a-c)** the bottom wire (indicated by the red box) is controllably toggled, while in **c-e)** the top wire (indicated by the green box) is controllably toggled without any effect on the neighboring wire.

STS measurements were performed on one end of a 10-dimer-long wire to study the transition between stationary and rapidly switching buckled dimers. To capture the rapid switching outcome more effectively than with a single scan, we also recorded the backward sweep to gain information on the final configuration of the wire. The results are shown in Figure 5. Four possible scenarios are illustrated to demonstrate that the outcome cannot be predicted once the bias goes above the threshold of switching out of control. Figure 5a shows the tip starting and ending on a raised DB; Figure 5b shows starting with a raised DB and ending with a lowered DB; Figure 5c shows starting and ending with a lowered DB; and Figure 5d shows starting with a lowered DB and ending with a raised DB. This procedure also illustrates that by simply monitoring the current after a pulse applied to the end dimer, the resulting buckled state of the wire can be determined, without the need for full imaging of the wire. Therefore, a pulse and monitoring current routine can be applied to deterministically set the buckled state as desired.

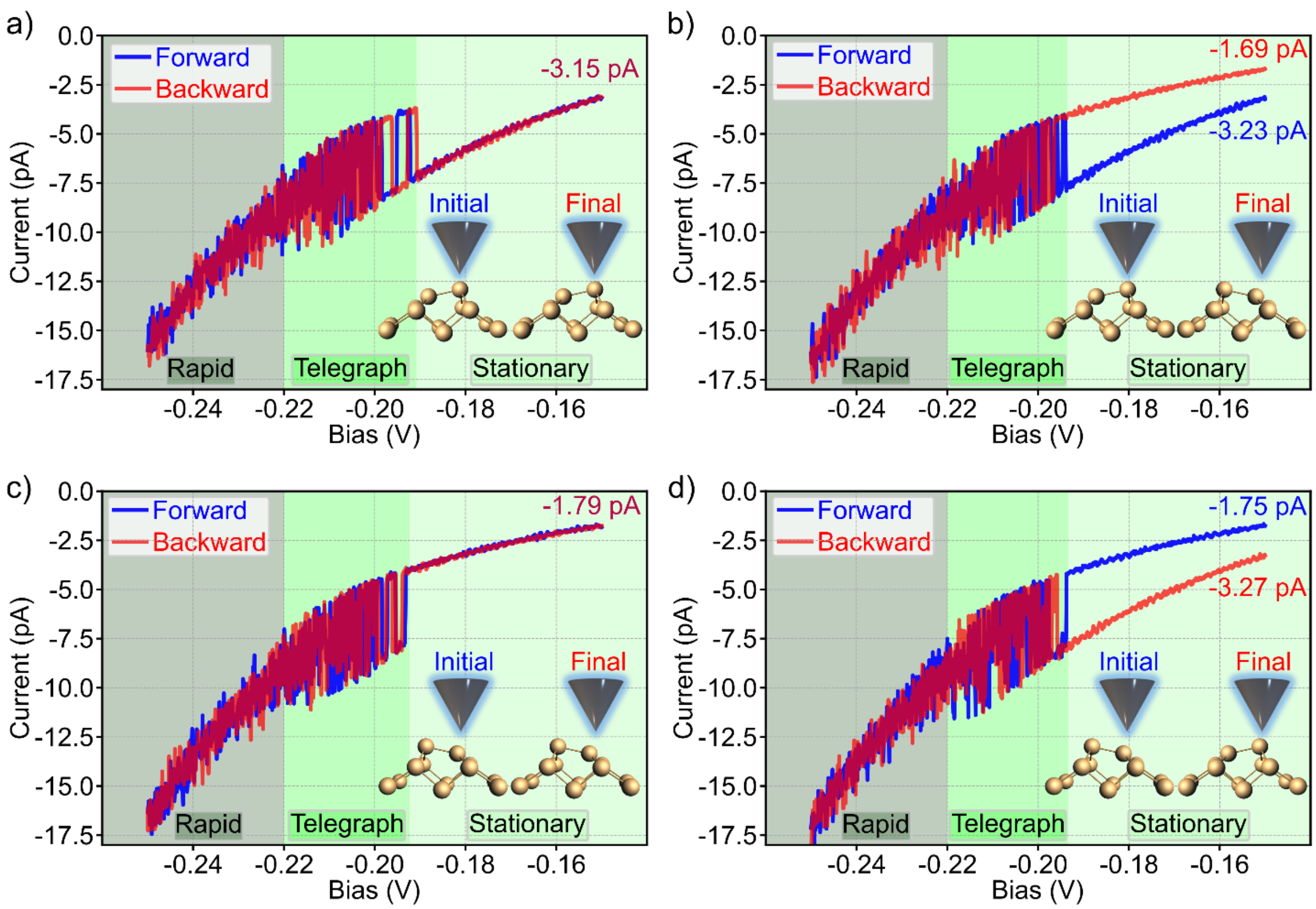

**Figure 5)** Series of I-V spectroscopy measurements showing four different outcomes after rapid shuffling of the buckled configuration of an end dimer in a 10-dimer long wire; **a)** Up-Up, **b)** Up-Down, **c)** Down-Down, **d)** Down-Up as the initial and final orientations, respectively. Each STS measurement starts at -150 mV and ends at -250 mV following the blue trace, then returns following the red trace. The plots illustrate stationary, telegraph, and rapidly switching regimes, represented with different shades of green in the background. The height setpoint was -200 mV and 2 pA on H-Si.

Panels in Figure 5 also illustrate the transition from the stationary regime to the rapidly switching regime, highlighting an intermediate phase characterized by telegraph noise, labeled as "Telegraph" in each panel. In this intermediate regime, the buckling orientation fluctuates uncontrollably but at a rate slow enough for the STM to capture, resulting in telegraph noise. This phenomenon is observed in Figures 1b and 1c, where STM images of three dimers show streaks caused by intermittent switching. In contrast, Figure 1a lacks streaks due to rapid switching, while Figure 1d shows no switching, resulting in a clear image. In the stationary regime, the current signal difference between the two buckled configurations is approximately 1.5 pA due to the very low bias. As the bias increases, this difference increases to ~5 pA in the telegraph regime and decreases to ~2 pA during rapid switching. The limited bandwidth (1 kHz) of the current amplifier prevents tracking of rapid configuration changes.

To investigate the telegraph noise in greater detail, we used a different setpoint on a single bare dimer to amplify the current difference between the two buckled configurations from ~5 pA to ~13 pA. After adjusting the new setpoint to -150 mV and 10 pA on a H-Si, the STM tip was positioned over one of the DBs of the dimer with the feedback disabled, thereby locking the tip-sample distance while the current signal was monitored. This was achieved by performing bias spectroscopy at the specified bias value over 250 seconds, with a 10 ms integration time, resulting in 25,000 data points (100 Hz signal acquisition rate), as shown in Figure 6a.

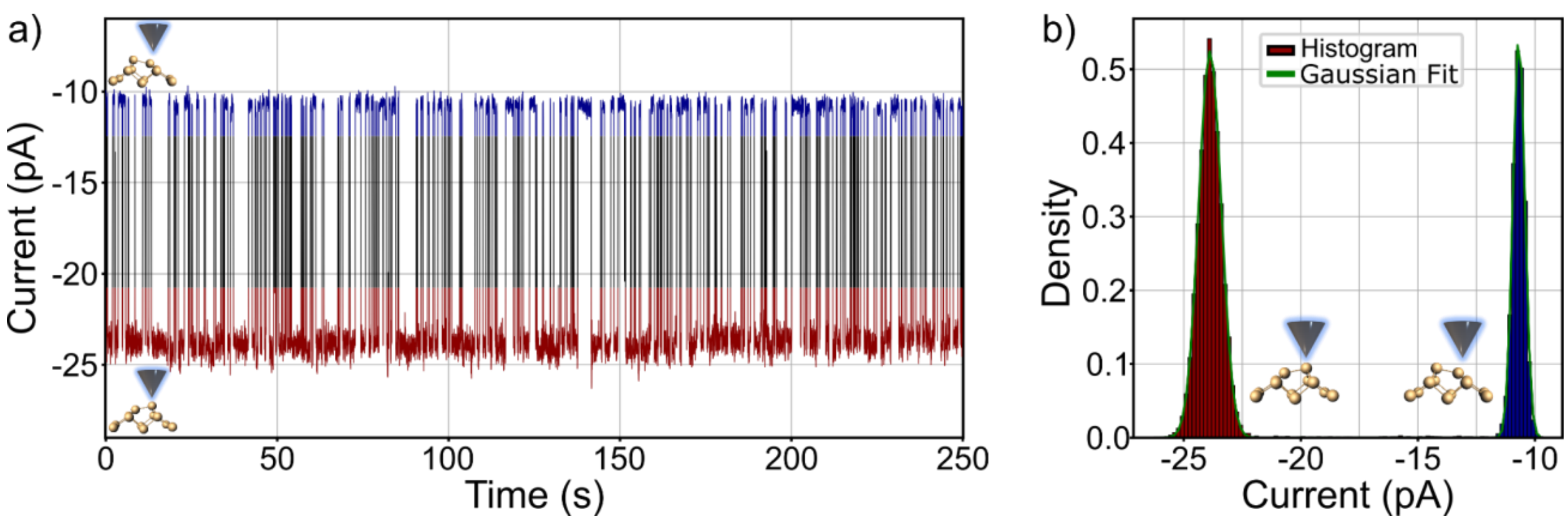

**Figure 6. a)** Telegraph noise of a single dimer. The upper part (blue) represents the DB in the lower position, while the lower part (red) shows the DB in the higher position. **b)** Normalized distribution of the two configurations, shown as a histogram representing the probability density of the two states. The height setpoint was -150 mV and 10 pA on H-Si.

Figure 6b illustrates the distribution of the data collected in Figure 6a, presented as a histogram. The two peaks observed in the histogram, representing the raised and lowered DB configurations, were fitted with a double Gaussian function. The goodness-of-fit tests resulted in Chi-squared and R-squared values of 0.093 and 0.999, respectively, indicating good agreement. This fit revealed that the two configurations are centered around -23.88 pA ($\mu_1$) and -10.69 pA ($\mu_2$), with standard deviations of 0.48 pA ($\sigma_1$) and 0.25 pA ($\sigma_2$), respectively. By considering $\pm 3\sigma$ for both peaks, the distribution shows that 63.6% of the data corresponds to the raised DB configuration, and 34.8% corresponds to the lowered DB configuration. The remaining 1.6% of the data lies between the two states, likely due to signal capture during transitions. This supports the predicted two-level nature of buckled dimers on the hydrogenated surface. While density functional theory-based potential energy surfaces or nudged elastic band calculations can predict the energy barrier between the two configurations, when it comes to experiment, this can be determined only via an Arrhenius analysis and would require additional data collected at different temperatures.

The uneven distribution observed in the histogram is attributed to perturbations caused by the presence of the STM tip, particularly due to attractive forces between the outermost tip atom and the raised silicon atom of the wire,[48, 49] favoring the raised position. In the absence of an STM tip, the telegraph noise is expected to exhibit a more even distribution between the two configurations within the limits allowed by the remaining inhomogeneities of the surface. Our proposed design for employing these wires without a probing tip includes samples with macro-to-atom connections. Communication to the atomic device that contains the wires will be combinations of binary wires and logic gates.[31]

## Conclusion

In conclusion, we have demonstrated that at low biases, bare silicon dimers on a hydrogenated surface exist in one of two buckled configurations. These orientations can be flipped using bias pulses applied with the STM tip. We presented various lengths of fabricated dimer wires locked into and switched between either buckled configuration. Through STS, we identified distinct regimes: stationary buckled dimers, telegraph noise, and rapid flipping. The telegraph noise confirms the presence of two distinct buckled configurations, as illustrated in the histogram in Figure 5b. As shown in a previous study,[31] DB-based OR logic gates and binary wires were shown to be steps toward binary atomic silicon logic (BASiL) circuitry. Combining buckled dimer wires as studied here with the expanding catalogue of BASiL components[50-52] could enable the reading and writing of dimer wire orientations without the need for an STM tip.

## Methods

**Measurements.** All STM measurements were performed using an Omicron low-temperature STM (LT-STM) at a temperature of 4.5 K and a pressure of ~$5\times10^{-11}$ Torr. Specific measurement parameters are provided in the text or figure captions. STM images presented in this study were analyzed using Gwyddion software.[53]

**Sample.** A degenerately doped p-type Si(100) sample was used throughout the study. The sample was cleaved from commercial wafers, degassed at 600 °C for approximately 18 hours, and then subjected to flash annealing at 1250 °C multiple times. After the final 1250 °C flash, the sample was terminated with a monolayer of hydrogen at 330 °C for 2 minutes, with hydrogen cracked from molecular hydrogen using a direct-current heated tungsten filament.[20]

**Tip.** The tungsten tip employed in this study was electrochemically etched using a 2M NaOH solution.[54] After etching, the tip was solvent-cleaned to remove byproducts and then heated to eliminate the oxide layer. It was subsequently sharpened to single-atom precision using a Field Ion Microscope.[55] In-situ tip shaping on Si DB patches was performed as needed.

## Acknowledgements

Authors would like to thank Ryan H. Foote for the useful discussions. Funding has been supplied by the National Research Council of Canada, Alberta Innovates Technology Futures, Natural Sciences and Engineering Research Council of Canada, and the Digital Research Alliance of Canada.

**Supplemental Information for:**

# Structural Control of Atomic Silicon Wires

Furkan M. Altincicek[1], Christopher C. Leon[1,4], Lucian Livadaru[1], Taras Chutora[1], Max Yuan[1], Roshan Achal[2], Jason Pitters[3], Robert Wolkow[1,2]*

[1] *Department of Physics, University of Alberta, Edmonton, Alberta T6G 2E1, Canada*

[2] *Quantum Silicon Inc., Edmonton, Alberta T6G 2M9, Canada*

[3] *National Research Council of Canada, Edmonton, Alberta T6G 2M9, Canada*

[4] *Current address: Département de chimie, Université Laval, Québec, Québec G1V 0A6, Canada*

* Correspondence to: rwolkow@ualberta.ca

## Electronic structure and NEB calculations by DFT

Geometry optimization calculations of dimer wire were performed by the Density Functional Theory (DFT) method using the AMS 2024 program[1-6] using a periodic supercell (slab) model consisting of 432 Si atoms in 6 layers with the top layer reconstructed in 2×1 (100) surface and H-capped and the bottom layer bulk terminated and H-capped as seen in Figure S1a. This supercell has 3 dimer rows at the surface, each consisting of 12 dimers. For an efficient calculation, gamma-only point was used for *k*-space and the supercell was divided in two regions to be treated with either a coarse or a fine accuracy, according to their importance in the present study. The first or "central" region consists of the atoms of the dimer row where the dimer wire is located, minus the end Si dimers adjacent to the border of the supercell. The remaining atoms were allocated to a "support" region. A double-zeta polarized basis set was used for the atoms in the central region, while just a double-zeta basis for the support region. We made use of a frozen core approximation for all atoms. The BAND DFT engine of AMS software was used for all calculations presented.

Calculations were performed with the GGA(PBE) exchange-correlation functional.[7] This DFT code uses localized basis-functions, of either Slater-type orbital (STO) or numerical atomic orbital (NAO) type, together with frozen core approximations of the inner electron shells of the atoms in the system. In our case, a double-zeta basis set was used for most of the atoms. However, we delimited a "central region" consisting of the clean dimer atoms and the 4 Si atoms directly bonded to it, where we used a superior double-zeta polarized basis set. This central region was also allowed flexibility during geometry optimization, while all other atoms were fixed. We performed geometry optimization for dimer wires of length 1, 2, 3, 5, 7 using for convergence criteria an energy threshold of 0.2 meV and gradient threshold of 0.02 eV/Å. In Figure S2, we give an example of geometry optimization results for a 3-dimer wire. We also checked that the ground state is doubly degenerate w.r.t. geometry – for each ground state geometry found, there is another configuration with the same energy and opposite dimer bond orientation.

The nudged elastic band (NEB) method[8] in DFT was used to determine the optimum path and transition state between the initial and final states. DFT calculations were performed using AMS 2024[1-6] with the PBE functional. In the NEB calculation we used 8 or 9 images, a spring constant of 5 eV/Å$^2$, and a force tolerance of 0.02 eV/Å. The climbing image NEB method was employed[9,10], starting from the initial NEB calculation. NEB calculations were performed on both a finite-sized nanocrystal model and an infinite chain (rod) model periodic in the *x*-direction. The Si nanocrystal model ($Si_{63}H_{56}$) has a 5-dimer wire and the surface and is otherwise H-capped all around. Calculations were performed with the GGA(PBE) exchange-correlation functional.[7] For checking the accuracy of the finite nanocluster model and for more extended dimer wires, we also created a periodic chain/ rod model, with a $Si_{140}H_{86}$ supercell spanning 10 dimers in the (periodic) *x*-direction and 6 Si layers in depth. The top and the bottom of the slab were H-capped, except for at the clean dimers of interest.

The geometry and total energy of initial, intermediate, and final states are obtained from each NEB calculation. The transition (barrier) state in each case is one in which the end dimer bond is flat (surface parallel). The molecular model in Figure S3 shows the barrier state for the 3-dimer wire (the other cases have very similar barrier states). The barrier height from the ground state (leftmost data point set as zero) is 270, 255, and 261 meV for the three cases, respectively. From the kink state, the barrier is 120, 130, 117 meV, respectively.

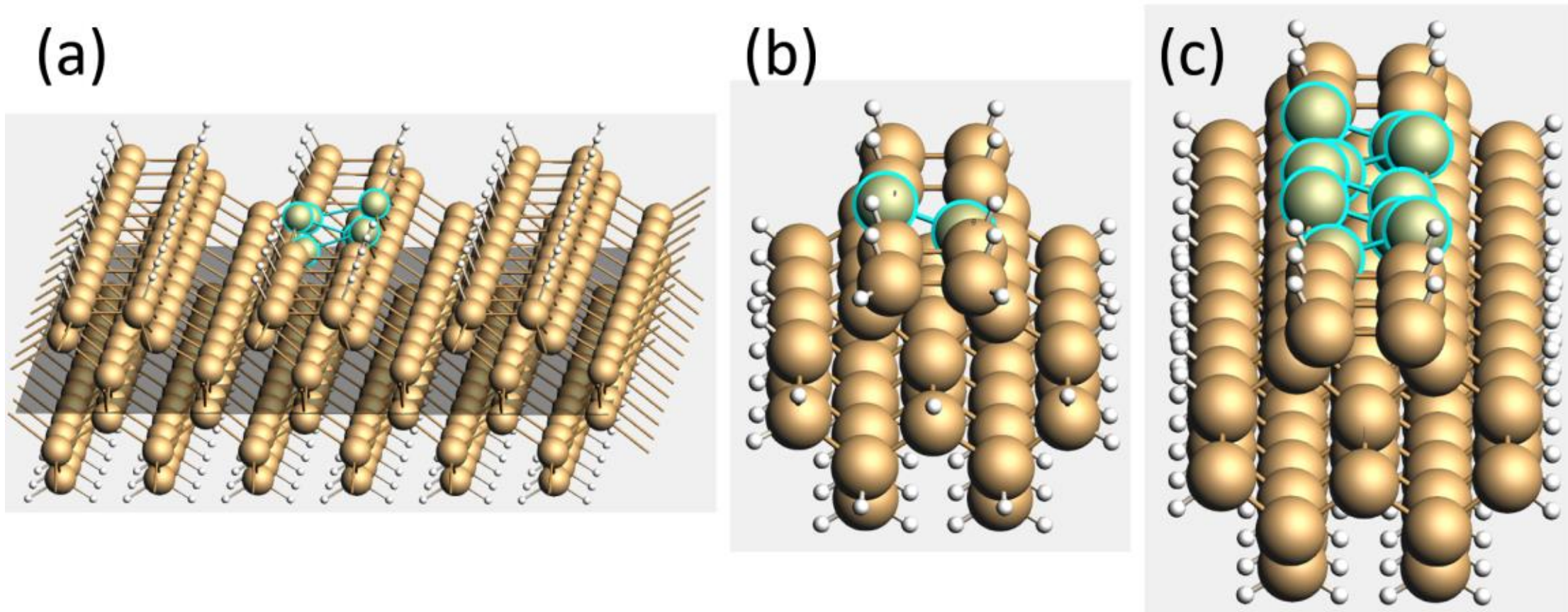

**Figure S1:** Molecular models used in our DFT calculations. (a) Periodic slab model, with a $Si_{432}H_{210}$ supercell with 12×3 surface unit cells, used in geometry optimization calculations. (b) Finite-sized nanocluster, $Si_{63}H_{56}$, with the dimer at the top used in NEB calculations. (c) Periodic rod model, with a $Si_{140}H_{86}$ supercell spanning a row with 10 dimers in the (periodic) *x*-direction used in NEB calculations. Silicon atoms are depicted in gold color, H atoms in white, and the Si atoms hosting DBs are highlighted in light green.

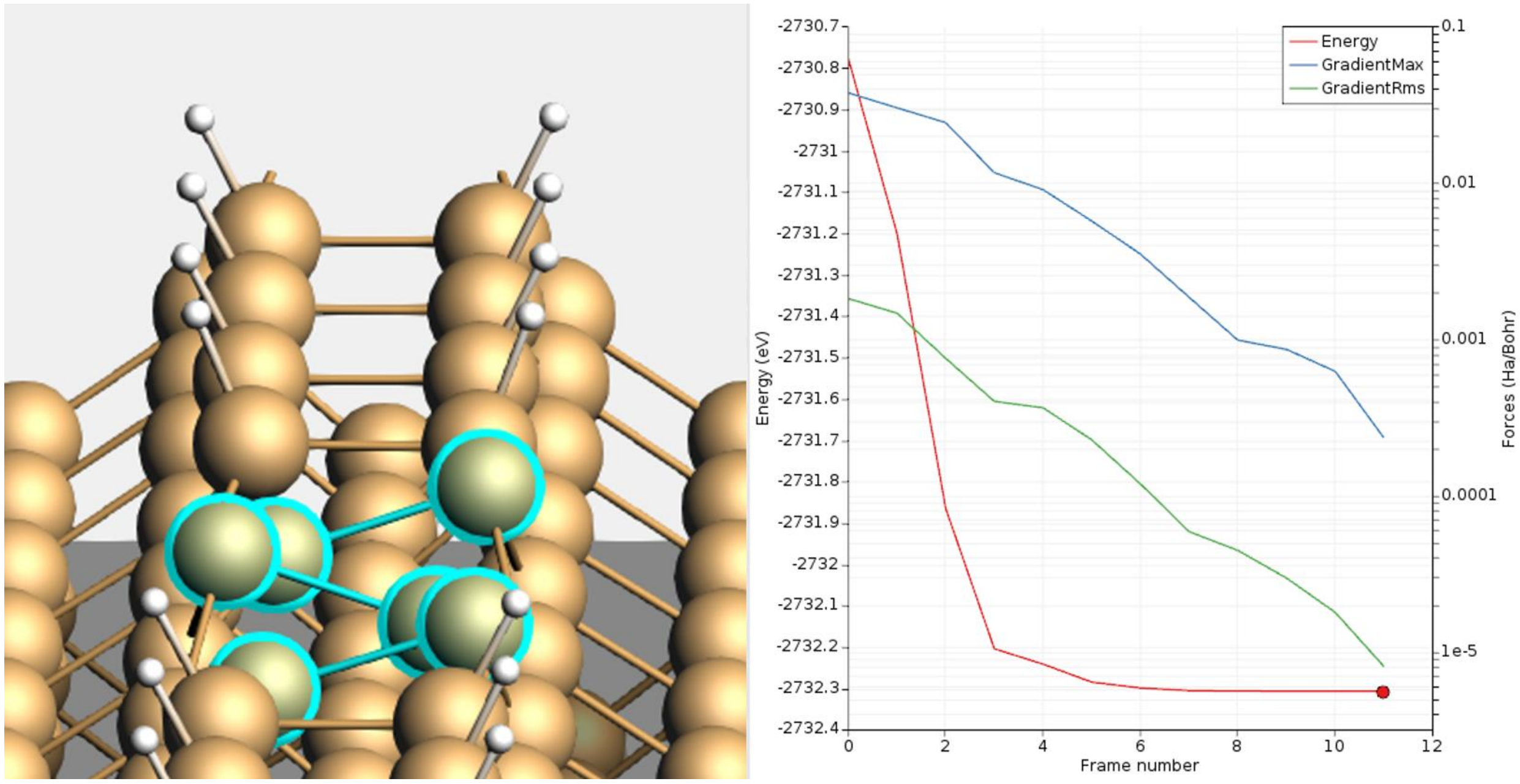

**Figure S2:** Result of a DFT geometry optimization calculation for a 3-dimer wire in a slab model. Energy threshold for convergence is 0.2 meV and the gradient threshold is 0.02 eV/Å. The final geometry is shown on the left and the evolution of energy and gradient is shown on the right.

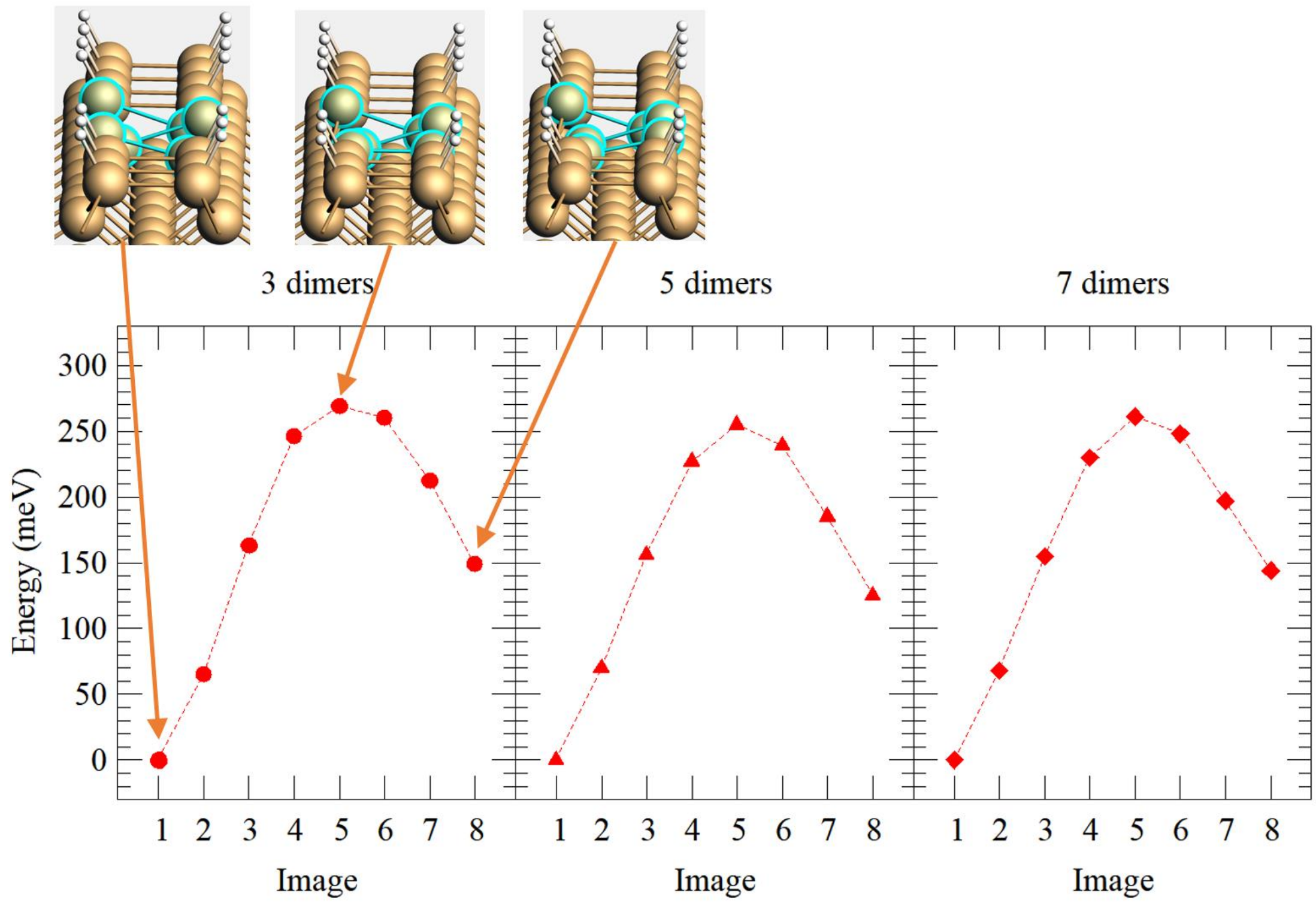

**Figure S3:** Results of DFT NEB calculations for the transition from the (initial) ground state of a dimer wire to a (final) metastable state with a kink at the very end of the wire. We show results for dimer wires of length 3, 5 and 7 on the rod model. In all cases, we used 8 images (including the ends), a spring constant of 5 eV/Å$^2$, and a force tolerance of 0.02 eV/Å. For the 3-dimer wire we also show the geometries of the initial, barrier and final states (images). The barrier height from the ground state (leftmost data point set as zero) is 270, 255, and 261 meV for the three cases, respectively. From the kink state, the barrier is 120, 130, 117 meV, respectively.

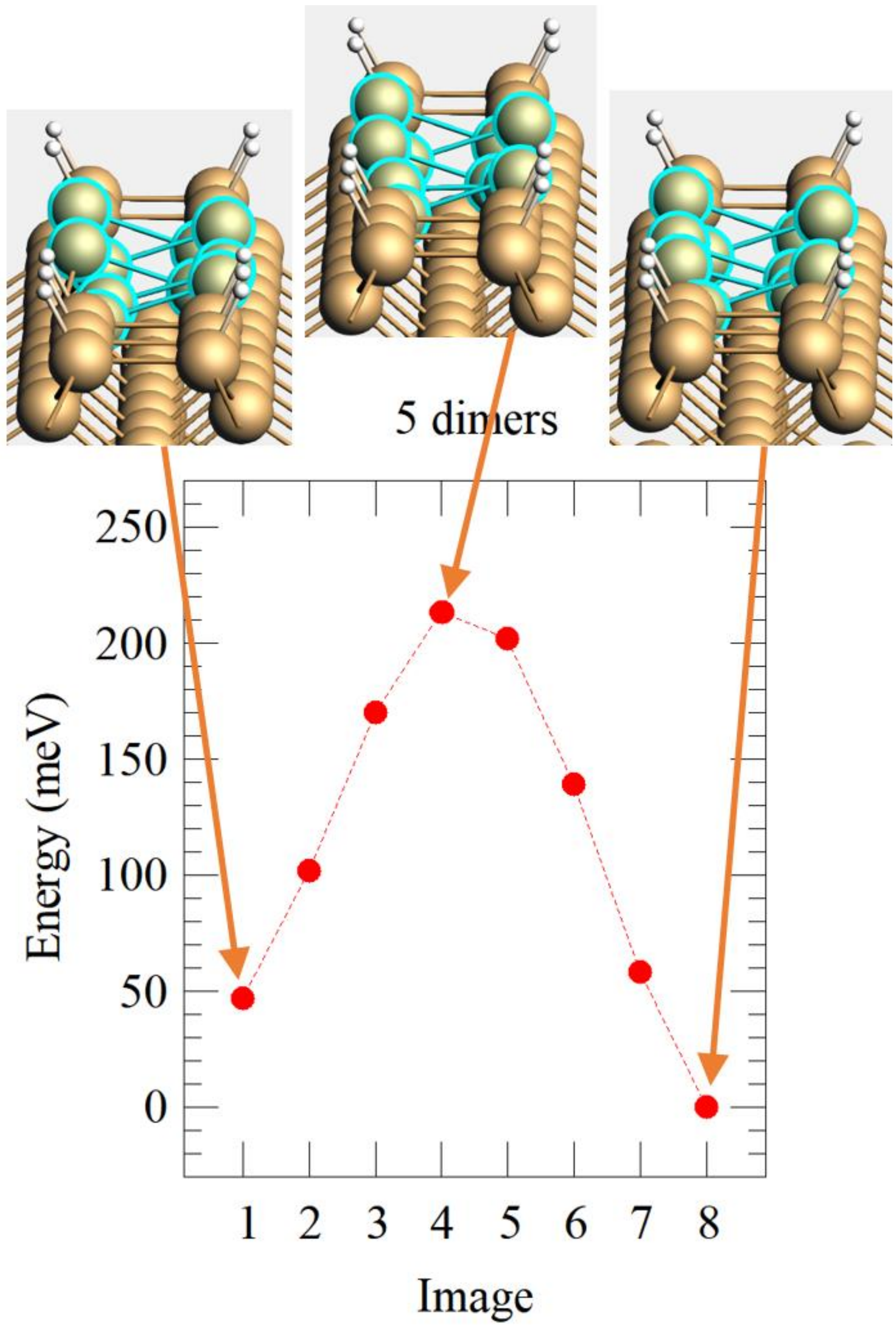

**Figure S4:** Results of DFT NEB calculations for the transition between two metastable states: one with a kink at the very end of the wire (image 1), the other with the kink displaced toward the interior by one dimer (image 8). We show results for dimer wires of length 5 on the rod model. We used 8 images (including the ends), a spring constant of 5 eV/Å$^2$, and a force tolerance of 0.02 eV/Å. We also show the geometries of the initial, barrier and final states (images).